\documentclass[11pt,a4paper]{article}
\usepackage{mathrsfs}
\usepackage{bm}
\usepackage{amsmath,amssymb}
\usepackage[colorlinks]{hyperref}%arXiv
\usepackage[dvipdfmx]{graphicx}
\usepackage[top=25truemm,bottom=25truemm,left=20truemm,right=20truemm]{geometry}
\makeatletter
	
		\@addtoreset{equation}{section}
\makeatother

\title{Tracking Pre-inflation Era \\ from density perturbation spectra}
\author{Tadashi Sasaki\footnote{t-sasaki@particle.sci.hokudai.ac.jp} \quad and \quad Hisao Suzuki\footnote{hsuzuki@particle.sci.hokudai.ac.jp} \\[5pt] \emph{Department of Physics, Hokkaido University, Sapporo 060-0810, Japan}}
\date{}
\newcommand{\teta}{{\tilde{\eta}}}
\newcommand{\tk}{\tilde{k}}
\newcommand{\apj}{Astrophys. J}
\begin{document}

\maketitle

\abstract{
One of the great triumphs of the inflationary model is the prediction of the flat power spectrum of the CMB fluctuation. 
The prediction is based on the assumption of the de-Sitter vacuum in the past infinity. However, the true past infinity of the inflation is expected to be dominated by radiation and curvature of the space. 
We consider pre-inflation era as dominated by radiation and curvatures as well as inflation potential. 
We derive the exact solutions for the scalar fields in this era and find a exact power spectra caused by the inflaton vacuum fluctuation. 
We show that the power spectrum is almost flat for sub-horizon scale and deviates from flat for very high super-horizon fluctuation, 
which is quite sensitive to the radiation and the curvature in the pre-inflation era.
}

%\begin{titlepage}
%\setcounter{page}{0}
%\begin{flushright}
%EPHOU 08-???\\
%February 2008\\
%\end{flushright}

\section{Introduction} 
  Inflationary scinario solves the problems of the flatness, horizon, and the origin of the fluctuation at the same time. 
One of the remarkable prediction of inflation is the power spectrum of the cosmic microwave back-ground (CMB). 
These prediction has been confirmed in great accuracy and provide severe constraints on the inflationary models\cite{Adeetal2016a}.
  
  However, it has been pointed out that there is a discrepancy of the Hubble parameters 
if we assume the flat-$\rm{\Lambda}$CDM model\cite{Sahnietal2014,Dingetal2015,Zhengetal2016}. 
Another problem which may be related with the Hubble constant is that
the CMB data favor the value $H_0 = (67.6 \pm 0.6) \rm{km s}^{-1}\rm{Mpc}^{-1}$\cite{Adeetal2016a} 
whereas the local measurement favors $H_0 = (73.24 \pm 1.74) \rm{km s}^{-1}\rm{Mpc}^{-1}$\cite{Riessetal}. 
Since we do not know the reason of this discrepancy, it may be wise to find other models.

  In the inflationary models, the seed of the fluctuation is the quantum theory of the inflaton. 
The quantum fluctuation of inflaton or some other scalar fields related to inflation (for example, the fields in the hybrid inflation) freezes to some classical value through the exponential expansion of the universe. 
In the theoretical side, it is known that there is an ambiguity in the choice of the vacuum. 
For example, in de-Sitter spacetime, it is known that there is no de-Sitter invariant vacuum so we have to choose a vacuum state. 
%One natural choice is that we choose a mode functions in such a way that they reduces to the ordinary modes in the past infinity. 
One natural choice is the vacuum state associated with the mode function which reduces to the ordinary positive frequency mode in Minkowski spacetime
in the past infinity.
Namely, the mode function $v_k$ is chosen by the requirement for the "in-state":
  \begin{equation}
  \lim_{\eta \rightarrow -\infty}v_{\omega}(\eta)=\frac{1}{\sqrt{2\omega}}e^{-i\omega \eta}
  \end{equation}
  This requirement fixes our vacuum state in de-Sitter expansion era. 
However if we have pre-inflation era, the past infinity of the inflation era may be affected by the history of the pre-inflation evolution.
Assuming that the inflation is well below the Planck scale, we expect the pre-inflation evolution of the universe is dominated by radiation as well as the effects of spatial curvature. 
Note that if our universe started with the quantum to classical transition, 
we naturally expect that the kinetic energy of the space is the same order as the potential energy caused by the spatial curvature, which is the origin of the flatness problem.

The effect of the spatial curvature for density perturbation in inflation has been analyzed in several papers \cite{RatraPeebles1994,RatraPeebles1995,Bucheretal1995,LythWoszczyna1995,Yamamotoetal1995}. 
They showed that the power spectrum for low $l$ region deviates from flat spectrum.
The analysis comparing to Plank results has been done in Ref.\cite{BGY, ORS}.
However, there is the  possibility that the radiation in the pre-inflation era also affects the choice of the vacuum. 
Usually we expect that they affects the overall normalization as a transfer
function.

In this paper, we consider the effect of both the radiation and the curvature, and will show the exact solution of the inflaton (massless scalar field) equation. 
By quantizing the inflaton, we will derive the exact power spectrum and show that it is almost flat expect for the very large scale. 
In the visible scale, the modification seems to be very small but may be observable.

In the next section, we will consider the pre-inflation era and will find that the scale factor is written by Weierstrass elliptic function in conformal time.
In section 3, we will show that the field equation of the inflaton can be written as Lam\'e equation so that the solutions can be written by elliptic functions. 
By using various formulas concerning elliptic functions, we derive the exact power spectrum.
The final section will be devoted to the discussions.

  %%%%%%%%%%%%%%%%%%%%%%%%%%%%%%%%%%%%
\section{Pre-inflation era}

We consider the Friedmann-Robertson-Walker metric in conformal time
\begin{equation}
	ds^2=a^2(\eta)\left[-d\eta^2+\frac{dr^2}{1-Kr^2}+r^2d^2\Omega \right].
\end{equation}
We do not know much about pre-inflation era. The inflation may start from quantum to classical transition in gravity. 
However, inflaton may be related to the Grand Unified Theories where the energy scale may be lower than the Planck scale. 
Therefore, it is reasonable to think that our universe began with many relativistic matters. 
If our universe started with the quantum to classical transition, it is reasonable to expect that the kinetic energy of our universe, 
which is related to expansion rate, and the potential energy, which is related to the spatial curvature, are the same order.  
The spatial curvature is suppressed during inflation to resolve the flatness problem.
In the inflationary scenario, the vacuum energy caused by the potential is the origin of the inflation era. 
There are many interpretation of the origin of the potential. 
In the chaotic inflation, stochastic process is the origin of the initial value of the potential. 
But here we assume that even before the inflation, the energy caused by the effective potential is present whose value is denoted by $V_0(>0)$.
Then the Friedmann equation is given by
\begin{equation}
	\left(\frac{1}{a^2}\frac{da}{d\eta}\right)^2=\frac{8\pi G}{3}\rho-\frac{K}{a^2},
\end{equation}
where $K$ is the curvature of the space and $a'=da/d\eta$. The energy density is dominated by the radiation. 
Therefore, we include the radiation density as well as the inflaton potential:
\begin{equation}
	\rho=\frac{\rho_r}{a^4}+V_0.
\end{equation}
We assume that quantum to classical transition occured at Plank scale and the curvature energy is almost the same order of energy of the radiation. 
Therefore, after the quantum to classical transition of the space-time, our universe is radiation and curvature dominant followed by the era of
vacuum energy dominant. Exponential growth starts when
\begin{equation}
	\frac{|K|}{a^2}\sim\frac{8\pi G}{3}V_0.
\end{equation}

We will use the following notation
\begin{equation}
	\frac{8\pi G V_0}{3}= H^2, \quad A=-\frac{K}{H^2},\quad B=\frac{\rho_r}{V_0}.
\end{equation}
Then we have 
\begin{equation}
	H\eta=\int^a\frac{da}{(a^4+Aa^2+B)^{1/2}}\label{eq:defeta}.
\end{equation}
$a^4+Aa^2+B=0$ has two solutions, $a^2=-A/2+\sqrt{(A/2)^2-B}$ and $a^2=-A/2-\sqrt{(A/2)^2-B}$, which will be denoted by $\tilde{e}_2,\tilde{e}_3$, respectively.
There are two cases: (i) $A>-2\sqrt{B}$, and (ii) $A<-2\sqrt{B}$.
%When both the roots are not positive real numbers, {\it i.e.} $A>0$ or $A^2<4B$, we can think that the universe started from the $a=0$ so that
For the case (i), the universe can start from $a=0$ since the singularities in the integrand of (\ref{eq:defeta}) are not on the real axis.  
Therefore we can fix the integration constant as 
\begin{equation}
	H\eta=\int_0^a\frac{da}{(a^4+Aa^2+B)^{1/2}}=\frac{1}{2}\int_0^{a^2}\frac{dx}{(x^3+Ax^2+Bx)^{1/2}},
\end{equation}
where $x=a^2$.
By shifting integration variable as $x=y-A/3$, we can remove the quadratic term,
\begin{equation}
	H \eta=\int_{e_1}^{a^2+e_1}\frac{dy}{[4(y-e_1)(y-e_2)(y-e_3)]^{1/2}},\label{yint}
\end{equation}
where 
\begin{equation}
	e_1=\frac{A}{3},\qquad e_2=\tilde{e}_2+\frac{A}{3},\qquad e_3=\tilde{e}_3+\frac{A}{3},
\end{equation}
which satisfy
\begin{equation}
	e_1+e_2+e_3=0.
\end{equation}
Since (\ref{yint}) can be decomposed as
\begin{equation}
	H\eta= \int_{e_1}^\infty\frac{dy}{[4(y-e_1)(y-e_2)(y-e_3)]^{1/2}}-\int_{a^2+e_1}^\infty\frac{dy}{[4(y-e_1)(y-e_2)(y-e_3)]^{1/2}},
\end{equation}
we can write the inverse relation by using the Weierstrass elliptic function as follows:
\begin{equation}
	a(\eta)=[\wp(\omega_1-\teta)-e_1]^{1/2},\label{eq:aone}
\end{equation}
where $\wp$ is defined as
\begin{equation}
	\wp(z)=\frac{1}{z^2}+\sum_{(m,n)\neq(0,0)}\left[\frac{1}{(z-2m\omega_1-2n\omega_2)^2}-\frac{1}{(2m\omega_1+2n\omega_2)^2}\right],
\end{equation}
and
\begin{equation}
	\teta=H\eta.
\end{equation}
$\omega_1$ is one of the half periods and given by
\begin{equation}
	\omega_1=\int_{e_1}^\infty\frac{dy}{[4(y-e_1)(y-e_2)(y-e_3)]^{1/2}}.
\end{equation}
It is easy to see that for small $\eta$ we have
\begin{equation}
	a(\eta)\sim \sqrt{\tilde{e}_2\tilde{e}_3}\teta=\sqrt{B}\teta,
\end{equation}
whereas $a$ approaches
\begin{equation}
	a(\eta)\sim \frac{1}{\omega_1-\teta},
\end{equation}
when $\eta \rightarrow \omega_1/H$, which represents de-Sitter phase in conformal time.

For the case (ii), 
the integrand of (\ref{eq:defeta}) has two singularities on the positive real axis at $a=\tilde{e}_2$ and $a=\tilde{e}_3(<\tilde{e}_2)$.
%When $A^2>4B$ and $A<0$ ($\Leftrightarrow e_2 >0$), It happens if the universe start with the negative curvature ($K<0$). 
Thus, the universe starts with a finite value $a=\sqrt{\tilde{e}_2}$ and our universe does not have the ``initial singularity''.
%\footnote{For the initial condition $a=0$, it will start to contract at $a=\sqrt{\tilde{e}_3}$ 
In this case we have
\begin{equation}
	a(\eta)=[\wp(\omega_2-\teta)-e_1]^{1/2}, \label{eq:atwo}
\end{equation}
where
\begin{equation}
	\omega_2=\int_{e_2}^\infty\frac{dy}{[4(y-e_1)(y-e_2)(y-e_3)]^{1/2}}
\end{equation}
is another half period.
Behavior around $\eta\sim0$ is different from that of case (i),
\begin{equation}
	a(\eta)\sim\sqrt{\tilde{e}_2}\left(1+\frac{\tilde{e}_2-\tilde{e}_3}{2}\teta^2\right),
\end{equation}
while de-Sitter phase appears around $\teta\sim\omega_2$,
\begin{equation}
	a(\eta)\sim\frac{1}{\omega_2-\teta}.
\end{equation}

%%%%%%%%%%%%%%%%%%%%%%5

\section{Exact solution of massless scalar fields and the spectrum of the density perturbation}
In this section, we solve the equation of massless scalar field exactly on the background spacetime derived in the previous section, 
and quantize it to calculate the power spectrum.
The equation of massless scalar field $\psi$ is given by
\begin{equation}
	\frac{\partial^2}{\partial \eta^2}\psi(x,\eta)+\frac{2}{a}\frac{da}{d\eta}\frac{\partial}{\partial \eta}\psi(x,\eta)-{\Delta}\psi(x,\eta)=0.
	\label{fieldequation}
\end{equation}
If we decompose the solution as $\psi(x,\eta)=\chi_k(\eta)\phi_k(x)$, where $\Delta\phi_k(x)=-k^2\phi_k(x)$,
the equation for $\chi_k(\eta)$ is
\begin{equation}
\frac{d^2}{d \eta^2}\chi_k+\frac{2}{a}\frac{da}{d\eta}\frac{d}{d \eta}\chi_k+k^2\chi_k=0.
\end{equation}
We shall use the variable $\teta=H\eta$ and write the above equation as
\begin{equation}
\chi^{\prime\prime}_k+2\frac{a^\prime}{a}\chi^\prime_k+\tk^2\chi_k=0,
\end{equation}
where the prime denotes the derivative with respect to $\teta$ and $\tk=k/H$.
By rescaling $\chi_k$ as
\begin{equation}
v_k=a\chi_k,
\end{equation}
we have the following equation for $v_k$:
\begin{equation}
v^{\prime\prime}_k+\left(-\frac{a^{\prime\prime}}{a}+\tk^2\right)v_k=0.
\end{equation}
Inserting (\ref{eq:aone}) for the case (i), we find
\begin{equation}
\frac{d^2}{d\teta^2}v_k=\left[2\wp(\omega_1-\teta)+e_1-\tk^2\right]v_k\label{eq:eqv}
\end{equation}
The equation for the case (ii) is quite similar. We obtain
%When $A<0$ and $A^2>4B$, the equation is quite similar. We obtain
\begin{equation}
\frac{d^2}{d\teta^2}v_k=\left[2\wp(\omega_2-\teta)+e_1-\tk^2\right]v_k.
\end{equation}

We observe that these equations are the Lam\'{e} equation 
\begin{equation}
\frac{d^2}{dx^2}y(x)=\left[l(l+1)\wp(x)+h\right]y(x),
\end{equation}
with $l=1, h=e_1-\tk^2$.
%Therefore, equation(\ref{eq:eqv}) corresponds to the Lam\'{e} equation for $l=1, h=e_1-\tk^2$.
The solution of the Lam\'{e} equation for $l=1$ is a classical result\cite{RLH}\footnote{The method used in Ref. \cite{RLH} has been applied
to the evolution equation for gravitational waves in Ref. \cite{Exact}}. For case (i), two independent solutions are
\begin{align}
v_k&=a_0\frac{\sigma(\omega_1-\teta+c)}{\sigma(\omega_1-\teta)\sigma(+c)}e^{-(\omega_1-\teta)\zeta(+c)},\notag\\
v_{-k}&=a_0\frac{\sigma(\omega_1-\teta-c)}{\sigma(\omega_1-\teta)\sigma(-c)}e^{-(\omega_1-\teta)\zeta(-c)},\label{eq:exactsolutions1}
\end{align}
where $a_0$ is the normalization constant, and $\zeta(z)$ and $\sigma(z)$ are defined as \cite{HTF}
%where(\cite{HTF})
\begin{align}
&\zeta(z)=\frac{1}{z}+\sum{}^\prime\left[\frac{1}{z-\omega}+\frac{1}{\omega}+\frac{z}{\omega^2}\right], \\
&\sigma(z)=z\prod{}^\prime\left[\left(1-\frac{z}{\omega}\right)\exp\left(\frac{z}{\omega}+\frac{1}{2}\frac{z^2}{\omega^2}\right)\right],
\end{align}
where $\sum'=\sum_{(m,n)\neq(0,0)}, \prod'=\prod_{(m,n)\neq(0,0)}$, and $\omega=2m\omega_1+2n\omega_2$.
They are related to the Weierstrass elliptic function as follows:
\begin{equation}
	\wp(z)=-\zeta'(z),\ \ \zeta(z)=\frac{\sigma'(z)}{\sigma(z)}.
\end{equation}
%satisfying
%\begin{equation}
%\wp(z)=-\zeta^\prime(z).
%\end{equation}
%The function $\sigma(z)$ is defined as
%\begin{equation}
%\sigma(z)=z\prod{}^\prime\lbrace(1-\frac{z}{\omega})\exp[\frac{z}{\omega}+\frac{1}{2}(\frac{z}{w})^2]\rbrace,
%\end{equation}
%which satisfies
%\begin{equation}
%\zeta(z)=\frac{\sigma^\prime(z)}{\sigma(z)}.
%\end{equation}
The value of $c$ is defined as
\begin{equation}
\wp(c)=e_1-\tk^2.\label{eq:cdef1}
\end{equation}
Expansion around $z=0$ can be derived as
\begin{align}
\wp(z)&=\frac{1}{z^2}+\cdots,\notag\\
\zeta(z)&=\frac{1}{z}+\cdots,\notag\\
\sigma(z)&=z+\cdots.\label{eq:aroudorigin}
\end{align}
We also point out that $\wp(z)$ is an even function whereas $\zeta(z),\sigma(z)$ are odd functions.
Although the equation (\ref{eq:cdef1}) determines $c$ only up to the periods of $\wp(z)$
\footnote{There is another ambiguity because $\wp(-c)=\wp(c)$. Replacing $c$ with $-c$ corresponds to the change $v_k\leftrightarrow v_{-k}$. 
We will fix this ambiguity later (see (\ref{wpprime})).},
the solutions (\ref{eq:exactsolutions1}) are not ambiguous because these are periodic functions with respect to $c$,
namely 
\begin{equation}
v_k(c+2\omega_i)=v_k(c)\label{eq:periodicc}.
\end{equation}
This result can be derived by using quasi-periodic properties
\begin{equation}
\zeta(z+2\omega_i)=\zeta(z)+2\eta_i,\qquad \sigma(z+2\omega_i)=-\sigma(z)\exp[2(z+\omega_i)\eta_i],
\end{equation}
where $\eta_i=\zeta(\omega_i)$.

For the case (ii), the solutions are obtained as
\begin{align}
v_k&=a_0\frac{\sigma(\omega_2-\teta+c)}{\sigma(\omega_2-\teta)\sigma(c)}e^{-(\omega_2-\teta)\zeta(c)},\notag\\
v_{-k}&=a_0\frac{\sigma(\omega_2-\teta-c)}{\sigma(\omega_2-\teta)\sigma(-c)}e^{(\omega_2-\teta)\zeta(c)},\label{eq:exactsolutions2}
\end{align}
which are also periodic with respect to $c$.

Let us next prove that $v_k$ and $v_{-k}$ are complex conjugates when $B>A^2/4$ (included in case (i)).
By (\ref{eq:cdef1}), we find 
\begin{align}
c&=\frac{1}{2}\int_{0}^\infty\frac{dx}{x^{1/2}(x^2+Ax+B)^{1/2}}+\frac{1}{2}\int_{-\tk^2}^0\frac{dx}{x^{1/2}(x^2+Ax+B)^{1/2}}\notag\\
&=\omega_1+\frac{1}{2}\int_{-\tk^2}^0\frac{dx}{x^{1/2}(x^2+Ax+B)^{1/2}}. \label{cint}
\end{align}
The second term is pure imaginary so that
\begin{equation}
(c-\omega_1)^*=-(c-\omega_1),
\end{equation}
which leads to
\begin{equation}
	c^*=2\omega_1-c\sim-c, \label{cstar}
\end{equation}
where ``$\sim$'' denotes the equivalence up to the periods.
Note that we also used the fact that $\omega_1$ is real for $B>A^2/4$.
Then,
it is straightforward to prove
\begin{equation}
v_k^*=v_{-k}, \ \ {}^\forall k>0
\end{equation}
This relation also holds for $A<-2\sqrt{B}$ (case (ii)).

In the case $A>2\sqrt{B}$, on the other hand, the relation between $v_k$ and $v_{-k}$ depends on $k$.
This is because complex conjugate of $c$ behaves differently from (\ref{cstar}) as follows:
\begin{equation}
	c^*\sim
	\begin{cases}
		c \ \ (\tk^4-A\tk^2+B<0) \\
		-c \ \ (\tk^4-A\tk^2+B>0).
	\end{cases}
\end{equation}
This difference can be seen from the second term in the equation (\ref{cint}), whose integrand becomes real near $x=-\tk^2$ leading to $c^*\sim c$.
%Let us demonstrate this relation for the case $A>0$, where $e_3<e_2<e_1$ holds.
%As in the previous case, $c$ is evaluated by 
%\begin{equation}
%	c(\tilde{k})=\int_{e_1-\tilde{k}^2}^\infty\frac{dy}{\sqrt{4(y-e_1)(y-e_2)(y-e_3)}}, \label{ck}
%\end{equation}
%where we denote explicitly the $\tilde{k}$ dependence of $c$.
%If we take the integration path on upper half plane, equation (\ref{ck}) is decomposed as follows:
%\begin{equation}
%	c(\tilde{k})=
%		\begin{cases}
%			\omega_1-i\int_{e_1-\tilde{k}^2}^{e_1}\frac{dt}{\sqrt{e(e_1-t)(t-e_2)(t-e_3)}}, \ \ (\tilde{k}^2<e_1-e_2) \\
%			\omega_1+i\int_{e_2}^{e_1}\frac{dt}{\sqrt{4(e_1-t)(t-e_2)(t-e_3)}}
%		\end{cases}
%\end{equation}
%In this case, \textcolor{red}{we have $c^*=c-2\omega_3+2\omega_2$}. 
It follows then $v_k$ is real for the wave number $k$ such that $c^*(k)\sim c(k)$,
\begin{equation}
	v_k^*=v_k, \ \ \tk^4-A\tk^2+B<0.
\end{equation}
%This result shows that the wave function $v_k$ is deformed in this region so that the in state associated with $v_k$ can no longer be regarded as a free theory.
This result shows that the wave function $v_k$ is deformed in this region so that $v_k$ can no longer be regarded as a mode function to quantize.
For this reason, we concentrate on the case $A<2\sqrt{B}$ in the rest of this paper.
%This result shows that the wave frequency is deformed because the in state cannot be regarded as a free theory. 
%\textcolor{red}{However, it is natural to}

We are going to find the the normalization of the solutions. 
We will use the following normalization:
\begin{equation}
v_k\frac{d}{d\eta}v_k^*-v_k^*\frac{d}{d\eta}v_k=i,\label{eq:normalizationcondition}
\end{equation}
This normalization is equivalent to considering $v_k\sim e^{-ik\eta}/\sqrt{2k}$ for massless scalar field in flat space.
By explicit evaluation of (\ref{eq:exactsolutions1}) and (\ref{eq:exactsolutions2}) using the following formulas\cite{HTF}, 
\begin{align}
\sigma(u-v)\sigma(u+v)&=-\sigma^2(u)\sigma^2(v)[\wp(u)-\wp(v)],\notag\\
\zeta(u+v)&=\zeta(u)+\zeta(v)+\frac{1}{2}\frac{\wp^\prime(u)-\wp^\prime(v)}{\wp(u)-\wp(v)},\label{eq:additionaltheorems}
\end{align}
we find 
\begin{equation}
	v_k\frac{dv^*_k}{d\eta}-v^*_k\frac{dv_k}{d\eta}=-a_0^2H\wp'(c).
\end{equation}
$\wp'(c)$ is determined up to sign by the differential equation $(\wp'(z))^2=4(\wp(z)-e_1)(\wp(z)-e_2)(\wp(z)-e_3)$ with the definition of $c$ (\ref{eq:cdef1}).
Here we take 
\begin{equation}
	\wp'(c)=-2i\tk\sqrt{\tk^4-A\tk^2+B} \label{wpprime}
\end{equation}
to ensure that $v_k$ represents the positive frequency mode around $\eta\sim0$ as will be shown in the 
next paragraph.
Then, the normalization condition (\ref{eq:normalizationcondition}) gives
\begin{equation}
	a_0=\frac{1}{\sqrt{2\tk H}(\tk^4-A\tk^2+B)^{1/4}}.
%a_0=\frac{H}{\sqrt{2}k^{3/2}(1-A\tk^{-2}+B\tk^{-4})^{1/4}}.
\end{equation}

Before considering the power spectrum, we must choose the vacuum state of the quantum field.
To do so,
we first derive the behavior of the mode function $v_k(\eta)$ in the past infinity. The past infinity corresponds to $\teta=0$.
We rewrite $v_k(\eta)$ as
\begin{align}
v_k(\eta)/v_k(0)&=\exp\left[\ln\sigma(\omega_1-\teta+c)-\ln\sigma(\omega_1+c)-(\ln\sigma(\omega_1-\teta)-\ln\sigma(\omega_1))+\teta\zeta(c)\right],\notag\\
&=\exp\left[\int_{\omega_1}^{\omega_1-\teta}(\zeta(x+c)-\zeta(x)-\zeta(c))dx\right]. \label{vk}
\end{align}
By using (\ref{eq:additionaltheorems}), we have
\begin{align}
v_k(\eta)/v_k(0)&=\exp\left[\frac{1}{2}\int_{\omega_1}^{\omega_1-\teta}\frac{\wp^\prime(x)-\wp^\prime(c)}{\wp(x)-\wp(c)}dx\right]\notag\\
&=\left(\frac{\wp(\omega_1-\teta)-\wp(c)}{\wp(\omega_1)-\wp(c)}\right)^{1/2}
	\exp\left[\frac{-\wp^\prime(c)}{2}\int_{\omega_1}^{\omega_1-\teta}\frac{dx}{\wp(x)-\wp(c)}\right]. \label{vk2}
\end{align}
We evaluate (\ref{vk2}) for $\teta<\omega_1$. Since $\wp(\omega_1-\teta)=e_1+\rm{O}(\teta^2)$,
we obtain 
\begin{equation}
	v_k(\eta)/v_k(0)\sim \exp\left[-i\tk(1-A/\tk^2+B/\tk^4)^{1/2}\teta\right] \label{pastbehavior}
%v_k(\eta)/v_k(0)\sim k^{-1}(\wp(x)-\wp(c))^{1/2}\exp[-ik(1-A/k^2+B/k^4)^{1/2}\teta]
\end{equation} for small $\teta$.
%This result shows that in the past these are not free theory except for large $k$, 
%where the effect of the cosmological acceleration becomes small. 
This result shows that, in the past, $v_k$ behaves as the mode function in the flat spacetime, i.e. $e^{-ik\eta}$,  only for large $k$ while the wave number is deformed
for small $k$. So we fix the mode function by considering large $k$ behavior. 
%We find
%\begin{equation}
%v_k(\eta)\sim e^{-ik\eta}
%\end{equation}

We expand the quantum field as
\begin{equation}
	\chi(\eta,x)=\frac{1}{a(\eta)}\sum_k\left(a_kv_k(\eta)\phi_k(x)+a^\dagger_kv_k^*(\eta)\phi^*_k(x)\right).
%\chi(\eta,x)\sim a_0\frac{1}{a}\sum_k(a_kv_k(\eta)+a_k^\dagger v_k^*(\eta))\phi_k(x)
\end{equation}
This estimate of the asymptotic behavior is consistent with the normalization by (\ref{eq:normalizationcondition}).
We are considering large $\tk$ region, where flat space approximation is valid, therefore
%We are considering the wave length (which is valid for large $k^2$ region which can be approximately represented by flat space. Therefore we have
\begin{equation}
\chi(\eta,x)=\frac{1}{a}\int\frac{d^3k}{(2\pi)^{3/2}}(a_{\bf{k}} v_k(\eta)e^{i\bf{k}\cdot\bf{x}}+a_{\bf{k}}^\dagger v_k^*(\eta)e^{-i\bf{k}\cdot
	\bf{x}})=\int\frac{d^3k}{(2\pi)^{3/2}}(a_{\bf{k}}\chi_k+a^\dagger_{\bf{k}}\chi^*_k).
\end{equation}
By using the explicit solutions (\ref{eq:exactsolutions1}) and (\ref{eq:exactsolutions2}), we get
\begin{equation}
	\chi^*_k\chi_k=\frac{v_k^*v_k}{a^2}=\frac{H^2}{2k^3(1-AH^2/k^2+BH^4/k^4)^{1/2}}\left(1+\frac{k^2}{H^2a^2}\right).
\end{equation}
After inflation $(a\gg1)$, this value is frozen to
\begin{equation}
%<0\vert\chi_k^*\chi_k\vert0>=\frac{v_k^*v_k}{a^2}=\frac{H^2}{2k^3(1-AH^2/k^2+BH^4/k^4)^{1/2}}
\chi_k^*\chi_k\to\frac{H^2}{2k^3(1-AH^2/k^2+BH^4/k^4)^{1/2}}.
\end{equation}
By the usual definition of the power spectrum
\begin{equation}
\mathcal{P}_\chi(k)=\frac{k^3}{2\pi^2}\chi_k^*\chi_k,
\end{equation}
we finally obtain the following power spectrum:
\begin{equation}
\mathcal{P}_\chi(k)=\left(\frac{H}{2\pi}\right)^2\frac{1}{(1-AH^2/k^2+BH^4/k^4)^{1/2}}.
\end{equation}

One of the prediction of this spectrum is that al large scale {\it i.e.} sufficiently small $k$, the perturbation spectrum goes to zero whereas it goes to constant value at large $k$.
Small $k$ behavior is understood from (\ref{pastbehavior}). If we introduce the effective wave number $q(k)=k(1-A/\tk^2+B/\tk^4)^{1/2}$ to write
$v_k(\eta)/v_k(0)\sim e^{-iq\eta}$, we can see that $q(k)$ has the minimum $q_{\rm{min}}=H\sqrt{2\sqrt{B}-A}$ at $\tk=B^{1/4}$.
Thus, the radiation energy is a kind of infrared cutoff. As a result, the vacuum expectation value of $\chi^2$, which is evaluated as the integral
of $\mathcal{P}_\chi(k)/k$, is IR convergent in contrast to the usual de-Sitter vacuum case.

\begin{figure}[h]
\begin{center}
\includegraphics[clip,width=10.0cm]{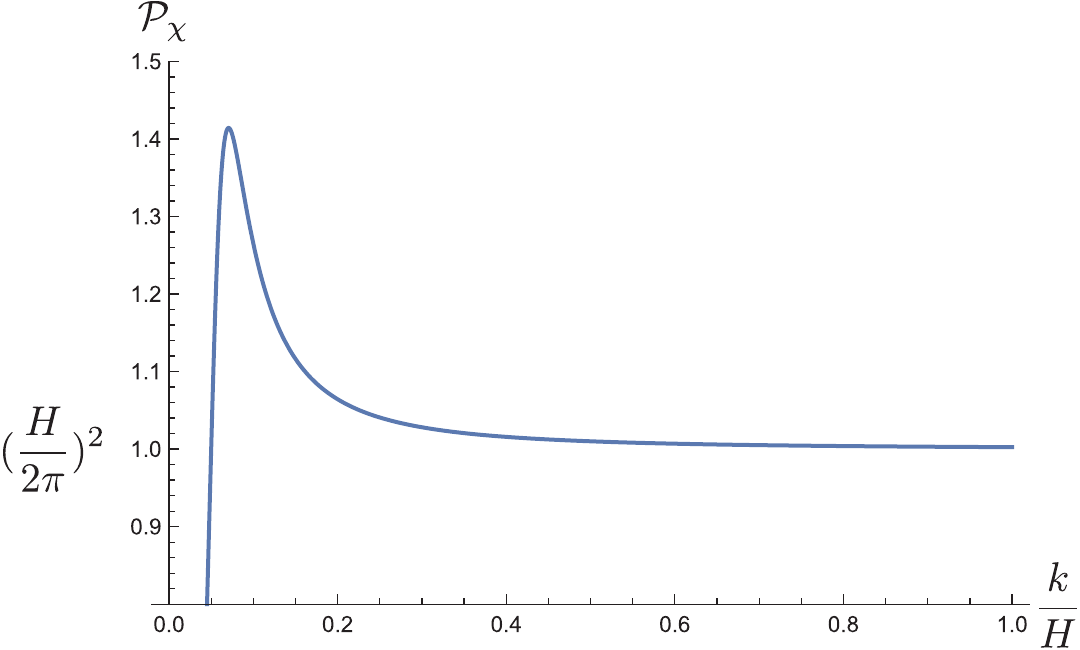}
\caption{A plot of the spectrum of $\mathcal{P}_\chi(k)$ normalized by $(H/2\pi)^2$ for open universe $A=5\times 10^{-3},B=2\times 5^4\times 10^{-8}$ at very high super horizon wavelength. We can see an enhancement of the power spectrum for small $k$.}
\label{fig:solutions1}
\end{center}
\end{figure}

\begin{figure}[h]%[htbp]
\begin{center}
\includegraphics[clip,width=10.0cm]{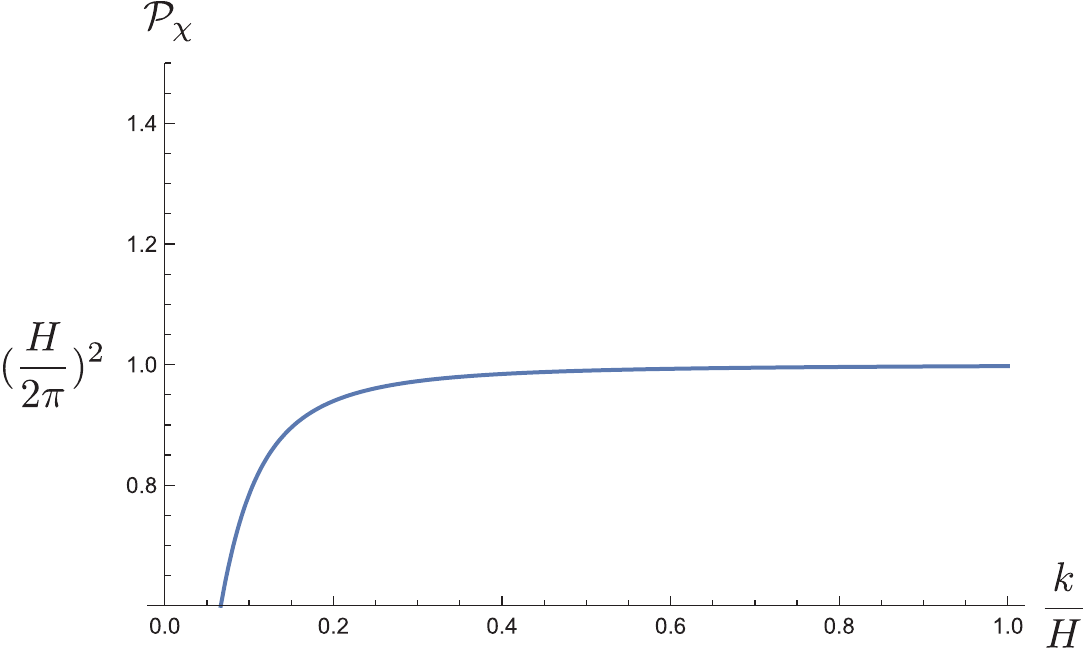}
\caption{A plot of the spectrum $\mathcal{P}_\chi(k)$ for closed universe $A=-5\times 10^{-3},B=2\times 5^4\times 10^{-8}$ at very high super horizon wavelength. We see no enhancement of power spectrum for small $k$.}
\label{fig:solutions2}
\end{center}
\end{figure}

When the curvature is negartive ($K<0 \Leftrightarrow A>0$), there appears an enhancement of the perturbation at small $k$.
As an example, we list a figure (Fig.\ref{fig:solutions1}) for open space $(K<0)$ for the values $A=5\times 10^{-3},B=2\times 5^4\times 10^{-8}$.
%When the curvature is positive, there is a enhancement of the perturbation when $k$ is small. For example, we list a figure \ref{fig:solutions1} for compact space $(K>0)$ at the values $A=-5\times 10^{-3},B=2\times 5^4\times 10^{-8}$.
For this parameter, we find that there is a very small deviation from flat space and there is a peak at $k=H\sqrt{2B/A}$, which may be invisible since the length scale is too large. 
However, if we consider closed universe ($K>0$), there is no enhancement but monotonically decrease as $k$ becomes smaller (Fig.\ref{fig:solutions2}). 
%If we consider open universe ($K<0$). there is no enhancement but decrease for small $k$. We show a fiture \ref{fig:solutions12}. 

%%%%%%%%%%%%%%%%%%%%%%%%%%%%%
\section{Summary and Discussions}
The usual inflationary scenarios assume that inflaton starts from the de-Sitter vacuum in the past infinity. 
We here considered that we have radiation and curvature dominant era before inflation. 
These stages affects the in-state vacuum compared with the case of usual inflation.
We have shown that the free scalar field equation (\ref{fieldequation}) in this scenario can be written as Lam\'e equation (\ref{eq:eqv}) and can be solved exactly. 
The solution can be written in terms of Weierstrass elliptic functions and we showed the exact power spectrum of the inflation. 
It modifies the usual scaling behavior, especially for small $k$. 
Although the effect of the modification seems very small, it is interesting that the scalar field equation can be written as Lam\'{e} equation and we could find the solution exactly. 

There are some problems, however. 
One is our assumption that the inflaton potential is present as constant even before inflation.
%We have treated the inflaton potential is present as constant even before inflation. 
There are many scenarios for inflation, in some of which the vacuum energy happens as phase transition. 
For such a case, we have to consider effective potential before inflation which may change in accordance with the energy scale. 
Another problem is that we do not know whether it is valid to use free inflaton before inflation. 
The interaction may change the behavior of the spectrum. However, it is still interesting that the free scalar field can be solved exactly.
It is also interesting that at sufficient value of $l$, the power spectum of CMB is almost constant but it has also enhancement for small $l$ and it looks like going to zero when $l$ is very small, although the error is still large enough.

%%%%%%%%%%%%%%%%%%%%%%%%%%%%%%%%%%%%%%%%%%%%%%%%%%%%%%%%%%%%%%%%%%%%%%%%

\end{document}